\documentstyle[prb,aps]{revtex}

\newcommand{\be}{\begin{equation}} \newcommand{\ee}{\end{equation}}
\newcommand{\ba}{\begin{eqnarray}} \newcommand{\ea}{\end{eqnarray}}
\newcommand{\bit}{\begin{itemize}} \newcommand{\eit}{\end{itemize}}
\newcommand{\ben}{\begin{enumerate}} \newcommand{\een}{\end{enumerate}}

\newcommand{\la}{\label} \newcommand{\dpa}{\partial}

\renewcommand{\Im}[1]{\mbox{\rm Im}\left(#1\right)}

\begin{document}

\title{General dispersion equation for oscillations and waves\\
       in non-collisional Maxwellian plasmas}
\author{V.~N.~Soshnikov
\thanks{Krasnodarskaya str., 51-2-168, Moscow 109559, Russia.}}%
\address{%
Plasma Physics Dept.,
All-Russian Institute of Scientific and Technical Information\\
of the Russian Academy of Sciences
(VINITI, Usievitcha 20, 125219 Moscow, Russia)}%
\maketitle
  \vspace{-0.15\textheight}

   \hspace*{0.5\textwidth} 
    \underline{\bf http://xxx.lanl.gov/e-print/physics/9706041}\\
    \vspace{0.13\textheight}

\begin{abstract}
  We propose a new and effective method to find plasma oscillatory
and wave modes. It implies searching {\em a pair of poles} of
two-dimensional (in coordinate $x$ and time $t$) Laplace transform
of self-consistent plasma electric field $E(x,t) \to E_{p_1p_2}$,
where $p_1\equiv -i\omega$, $p_2\equiv ik$ are Laplace transform
parameters, that is determining {\em a pair of zeros} of the
following equation $$\frac1{E_{p_1p_2}}=0 .$$ 
This kind of conditional equation for searching double poles of
$E_{p_1p_2}$ (which correspond to terms of the type 
$\frac{E_{11}}{\left(p_1-p_1^{(n)}\right)\left(p_2-p_2^{(n)}\right)}$
in double Laurent expansion of $E_{p_1p_2}$) we call ``general 
dispersion equation'', so far as it is used to find the pair values
($\omega^{(n)}, k^{(n)}$), $n=1, 2, \ldots$. 
It differs basically from the classic dispersion equation 
$\epsilon_l\left(\omega,k\right)=0$ (and is not its generalization),
where $\epsilon_l$ is longitudinal dielectric susceptibility,
its analytical formula being derived according to Landau analytical
continuation. 
In distinction to $\epsilon_l$, which is completely plasma 
characteristic, the function $E_{p_1p_2}$ is defined by initial and
boundary conditions and allows one to find all the variety of 
asymptotical plasma modes for each concrete plasma problem.
In this paper we demonstrate some possibilities of applying this 
method to the simplest cases of collisionless ion-electron plasma 
and to electron plasma with collisions described by 
a collision-relaxation term $-\nu f^{(1)}$.
\end{abstract}
\vspace{0.7mm}

\vspace{0.7mm}


\section{Introduction}

  Up to date the textbook approach to determining
dispersion dependence of oscillation frequency on wave number
$\omega(k)$ implies solving some dispersion equation,
the latter being found using Landau rule of bypassing poles
in his theory of collisionless damping (``Landau damping'',
see~\cite{bib-1}).
This theory is based on analytical continuation
of complex function defined as Cauchy integral
\be
 F(\omega) = \int\limits_{-\infty}^{\infty} 
                 \frac{\phi(z) dz}{z-\omega} \ ,
\ee
where the integration is performed primarily along the real $z$-axis. 
This integral is well defined for $\Im{\omega}>0$ and 
$\Im{\omega}<0$. 
Landau considers as physically defined just the function 
$F^+(\omega)$ in the upper half-plane of $\omega$ 
due to adiabatic in time switching on.
Then he tries to determine analytical continuation of this
function to the region $\Im{\omega}<0$ 
through the point $\Im{\omega}=0$ on the real axis, 
with deformation of the integration contour
to the lower half-plane.

  One could say that analytical continuation through 
integration contour $A$ (i.e. the real axis ${\bf R}$) 
does not exist by its very definition 
as a {\em regular} function, 
since $F^+_A(\omega) \equiv F^+(\omega)$ is indefinite 
and not regular 
for $\omega \in A$, i.e. for $\omega$ on the real axis, 
discriminated by physics of the problem. 
Suppose the fixed integration contour $A'$ is parallel to 
and lies above the contour $A$. 
When $\omega$ lies above $A'$ according to Cauchy theorem
we have $F^+_A(\omega)=F_{A'}(\omega)$.
But $F_{A'}(\omega)$ is indefinite for $\omega\in A'$ and 
therefore both functions $F^+_A$ and $F_{A'}$
for the first sight {\em cannot be regarded as analytical
continuation of one to another}.

  Nevertheless one can resolve above-mentioned paradox
and construct the analytical continuation by the following 
way.
Let contour $A'$ to lie over $A$ by $ib$, $b>0$.
According to Cauchy theorem for $\Im{\omega}>b$
we have $F_{A'}(\omega)=F^+_A(\omega)$,
so
\be
 \int\limits_{-\infty}^{\infty} 
  \frac{\phi(z) dz}{z-\omega} =
   \int\limits_{-\infty}^{\infty} 
    \frac{\phi(z+ib) dz}{z+ib-\omega}\ , 
\la{eq-b}
\ee
where $\phi(z)$ is supposed to be analytical 
in the stripe $0\le\Im{z}<b$. 
Analytical continuation of $F_{A'}(\omega)$ into the region 
$0<\Im{\omega}<b$ can be obtained by Eq.(\ref{eq-b}) 
with some arbitrary $b<\Im{\omega}$. 
In the same way if $\Im{\omega}<0$ one should use 
Eq.(\ref{eq-b}) with some arbitrary $b<\Im{\omega}$. 
In all cases we obtain the single analytic function
(due to singleness theorem).\footnote{%
For the analytical continuation of $F^+(\omega)$ 
from the upper half-plane (this is marked by 'plus'-sign
in contrast to analytical continuation of $F^-(\omega)$ 
from the lower half-plane, which is marked by 'minus'-sign)
to the lower one we have
$$ F^+(\omega) = {\cal PV}\int\limits_{-\infty}^{\infty}
      \frac{\phi(z) dz}{z-\omega} + i\pi\phi(\omega)
       \quad (\mbox{at}\ \Im{\omega}=0) ;$$
$$ F^+(\omega) = {\cal PV}\int\limits_{-\infty}^{\infty}
      \frac{\phi(z) dz}{z-\omega} + 2i\pi\phi(\omega)
       \quad (\mbox{at}\ \Im{\omega}<0) ;$$
$$ F^-(\omega) = {\cal PV}\int\limits_{-\infty}^{\infty}
      \frac{\phi(z) dz}{z-\omega} 
       \quad (\mbox{at}\ \Im{\omega}<0) ;$$
$$ F^+(\omega) = F^-(\omega) + 2i\pi\phi(\omega)
       \quad (\mbox{for all}\ \omega) .$$
Here $F^+(\omega)$ is regular at all $\omega$ with $
-b_0<\Im{\omega}\le 0$,
where $b_0$ is defined by the analiticity properties of $\phi(z)$.
Cf. also the analogous expressions in~\cite{bib-7,bib-8}.
These relations and all features can easily be demonstrated
with a simple example of the function 
$\phi(z)=1/\left(z^2+\omega^2\right)$ (noted by A.~P.~Bakulev)}
This function is obtained by deforming integration contour
in such a way that point $\omega$ lies always above the contour
(this is due to the $^+$-type of considered analytical function).
For Maxwellian plasma  
$\phi(z)\sim z\exp\left(-\alpha z^2\right)$ 
with $|\phi(z)|\to\infty$ at $\Im{z}\to\pm\infty$ and 
it has no poles in the complex plane $z\in C$.
Therefore the corresponding functions $F^{\pm}(\omega)$
are analytic functions in the whole complex plane.

Using analytical continuation of $F^+(\omega)$
in the lower half-plane (instead of $F^-(\omega)$)
is closely related with the absence of any solutions
of dispersion equation for any $\delta$
in $\omega=\omega_0\pm i\delta$,  $\omega_0 \neq 0$,
when dispersion equation is derived by substitution
of plane wave solution $\exp(-i\omega t + ikx)$
into Vlasov equations for electrons
(that is linearized kinetic equation + Poisson equation)
in the case of background Maxwellian plasma~\cite{bib-4}.
The theory of Landau damping can also be considered~\cite{bib-4}
as an equivalent to solving the analytically continued
to the region $\Im{\omega}<0$ dispersion equation,
that is equation $F^+(\omega,\vec k)=1$ with
analytically continued left-hand-side function 
$F^+(\omega,\vec k)$.

The {\em mathematical} part of Landau theory 
(see f.e.~\cite{bib-7}) 
contrary to the said in~\cite{bib-3,bib-4}
does not give rise to any objections.
However there are still the following {\em logical} 
objections:
\ben
 \item
  Landau solution in the form of a damping wave 
  {\em must satisfy} primary Vlasov equations, but it does not.
 \item
  In the original Landau contour integral $F(\omega)$ 
  there is the physically discriminated {\em real} axis
  of integration over velocity component $v_x$.
  One solves some initial problem in the time interval
  $0\le t<\infty$, and coordinate dependence is introduced
  through the common factor $\exp(ikx)$ for all times $t$.
  Here initial conditions really are in no connection with
  unknown conditions at $t<0$. Writing solution in the form
  of inverse Laplace transform (``original'') with parameter
  $p=-i\omega$ physically leads one to search in the first place
  some finite (damping or oscillatory non-damping) solutions
  with $\omega=\omega_0-i\delta$, $\omega_0>0$ and $\delta>0$.
  These solutions correspond to the lower half-plane of
  complex frequency $\omega$.\\
  The function $F^-(\omega)$ is analytical in the lower
  half-plane, and {\em there is no logical necessity to use
  analytical continued from the upper half-plane function}
  $F^+(\omega)$, or consider some analytical continuation into
  upper half-plane, where oscillatory solutions are exponentially 
  divergent. As it is also well-known, the integrand of inverse
  Laplace transform $L(F)$ in this case has no poles and,
  correspondingly, no solutions in the form of damping wave
  (that means the absence of solutions of equivalent 
  dispersion equation). 
 \item
    The application of two-dimensional Laplace 
  transformation~\cite{{bib-3},{bib-4}}
  (see also this paper) shows 
  that the asymptotic solution of Vlasov equations with
  Maxwell distribution function exists only in the form
  of a system of coupled oscillatory modes.
  The absence of solutions in the form of a single wave
  $\exp\left(-i\omega t-i\vec k\vec x\right)$, 
  where $\omega$ is complex value,
  can be proved by {\em reductio ad absurdum}:
  the substitution of this solution into Vlasov equations leads
  to a dispersion equation which has no solutions of the type
  $\omega=f\left(\vec k\right)$.
  Landau theory for the solution in a more general form
  $f(t)\cdot\exp\left(i\vec k\vec x\right)$ ought to be considered
  as a part of the classical proof by {\em reductio ad absurdum}:
  using one-dimensional Laplace transformation for the half-plane 
  of damping waves $\Im{\omega}<0$ one obtains, 
  contrary to initial supposition, that damping solutions in 
  the form $f(t)\sim\exp\left(-i\omega t\right)$ are absent,
  and moreover the analytical continuation into the half-plane
  $\Im{\omega}>0$ leads, also contrary to initial supposition,
  to exponentially growing solutions.
  Besides that the latters satisfy not the Vlasov equations 
  but some other equations with another dispersion relation,
  that proves in its turn the failure of the initial supposition
  about the assumed form of solution.
\een
  Therefore there is discrepancy between mathematical
correctness of Landau theory and unjustifyness and
unnecessity of its real application. 

   As it has been pointed out in~\cite{bib-2} the solution of 
``Landau problem'' is really non-damping standing oscillations 
(but not a travelling damping wave) as it has been proved 
in~\cite{bib-3,bib-4}, and the desired damping solution 
can be attained through the transition to 
a non-Maxwellian electron velocity distribution function 
(``cut-off'' velocity distribution).
Probably, it is equivalent to solving dispersion equation
which can be solved with respect to $\delta$ owing to
using just non-Maxwellian background distribution function.
But evidently, such solution is not completely defined
due to the arbitrariness of Maxwell distribution ``cut-off''.
And besides that, under special condition of ``cut-off''
$v<c$ where $c$ is the light velocity,
one should in any case solve the relativistic kinetic equation.

   A quite different, but general way to consider plasma
oscillations in frames of linearized Vlasov equations
was proposed in~\cite{{bib-3},{bib-4}}.
It relies on two-dimensional (in coordinate $x$ and time $t$)
Laplace transforms $f^{(1)}_{p_1p_2}$, $E_{p_1p_2}$
for the perturbation $f^{(1)}(x,t,\vec v)$ of distribution function
and the field $E(x,t)$, where $p_1$, $p_2$ -- are the parameters
$p_1=-i\omega$, $p_2=ik$ of Laplace transformation.
Depending on initial and boundary conditions
one obtains different analytical expressions for the function
$E_{p_1p_2}$. 
The equation for double poles (in $p_1$ and $p_2$)
of this function defines different asymptotic oscillatory modes
as the pairs ($\omega^{(n)}$,$k^{(n)}$), $n=1, 2, \ldots$\ .
So, it is natural to consider the equation for poles
of $E_{p_1p_2}$ as ``general dispersion equation''.
The general asymptotic solution is obtained as a sum of 
exponential modes with coefficients defined by residues, 
but substitution of this solution 
$f^{(1)}\left(x,t,\vec v\right)$ and $E(x,t)$
into Vlasov equations does not lead, in general, to any
dispersion equation connecting $\omega$ and $k$.
That is, the dispersion equation, in common sense, does not exist.
Indeed, it ought to suppose that, vice versa, for a given 
concrete problem its asymptotical solution in the form of coupled
oscillatory modes must identically satisfy 
Eqns.(\ref{eq-1a})-(\ref{eq-2}), and this can be tested
by direct substitution of solution in 
Eqns.(\ref{eq-1a})-(\ref{eq-2}).

  This approach appears to be more general,
allowing one to find additional oscillatory modes also
in the case of non-Maxwellian distribution functions,
when the usual dispersion equation could take place 
and have solutions.

  These statements which were developed in~\cite{bib-4}
appear to form a new and unexpected, though very simple,
approach to the problem of plasma oscillations.
In this paper we demonstrate the very possibility of
applying the two-dimensional Laplace transformation for finding
ion-electron oscillations and waves in collisionless Maxwellian
plasma and electron oscillations in low-collisional Maxwellian
plasma in the case of electron-neutral collisions described
by the collision-relaxation term $-\nu f^{(1)}$,
where $\nu \simeq \mbox{const}$ is collision frequency.
(But, if Landau theory is wrong, then it is necessary also
to revise collision corrections for accurate taking into account
the Coulomb collisions including one
obtained in ~\cite{bib-6} with the method of expansion
into asymptotically divergent series over small parameter
$\frac{\delta}{\omega_0}$.)

  To demonstrate our approach we consider here the simplest cases
of one-dimensional (in $x$) plane longitudinal plasma waves
based on kinetic equations for electrons, correspondingly, ions
\ba
  \frac{\dpa f^{(1e)}}{\dpa t}
  + v_x \frac{\dpa f^{(1e)}}{\dpa x}
  - \frac{|e|E_x}{m}\frac{\dpa f^{e}_0}{\dpa v_x}
  &=& 0
\label{eq-1a}\\
  \frac{\dpa f^{(1i)}}{\dpa t}
  + v_x \frac{\dpa f^{(1i)}}{\dpa x}
  + \frac{|e|E_x}{m}\frac{\dpa f^{i}_0}{\dpa v_x}
  &=& 0
\label{eq-1b}
\ea
and Poisson equation
\be
  \frac{\dpa E_x}{\dpa x} =
  -4\pi|e|n_e\int\limits_{-\infty}^{\infty}
     f^{(1e)} d^3\vec v
  +4\pi|e|n_i\int\limits_{-\infty}^{\infty}
     f^{(1i)} d^3\vec v; \quad n_e \simeq n_i. 
\label{eq-2}
\ee

\section{The boundary problem of exciting
         ion-electron oscillations and waves}

   The plasma is assumed Maxwellian, homogeneous and
infinite in space and time.
The boundary condition is assumed to be given in the plane $x=0$
(the plane geometry).
The one-sided Laplace transformation allows one to obtain
absolutely definitive solution in the region $0\le x<\infty$.
But for the same boundary condition one can also obtain
one-sided solution with the help of analogous Laplace transformation
in the region $-\infty<x\le 0$ (after substitution $x'=-x$).
So one obtains the united solution in the whole interval
$-\infty<x<\infty$.
The solution will be continuous at $x=0$.
The derivative $\dpa f^{(1)}/\dpa x$ at $x=0$ can be
either continuous or discontinuous.
The same is true also for the one-sided Laplace time transformation
at $-\infty<t<\infty$ for some given initial condition at $t=0$.
The united solution will not be specially analyzed here.

  Applying two-dimensional Laplace transformation to
perturbations of the electron and ion distribution functions
$f^{(1e)}$, $f^{(1i)}$
\ba
 f^{(1)}\left(x,t,\vec v\right)
 &=&\frac{1}{\left(2\pi i\right)^2}
     \int\limits_{\sigma_1-i\infty}^{\sigma_1+i\infty}
      \int\limits_{\sigma_2-i\infty}^{\sigma_2+i\infty}
       f^{(1)}_{p_1p_2} e^{p_1 t + p_2 x} dp_1 dp_2 \ ,
\label{eq-3}\\
 \frac{\dpa f^{(1)}\left(x,t,\vec v\right)}{\dpa x}
 &=&\ \frac{1}{\left(2\pi i\right)^2}
     \int\limits_{\sigma_1-i\infty}^{\sigma_1+i\infty}
      \int\limits_{\sigma_2-i\infty}^{\sigma_2+i\infty}
       p_2 f^{(1)}_{p_1p_2} e^{p_1 t + p_2 x} dp_1 dp_2
  - \frac{1}{2\pi i} f^{(1)}\left(0,t,\vec v\right)
     \int\limits_{\sigma_2-i\infty}^{\sigma_2+i\infty}
       e^{p_2 x} dp_2 \ ,
\label{eq-4}
\ea
and analogous expressions for
$\frac{\dpa f^{(1)}\left(x,t,\vec v\right)}{\dpa t}$,
$E(x,t)$, $\frac{\dpa E(x,t)}{\dpa x}$ in 
eqns.(\ref{eq-1a})--(\ref{eq-2});
assuming for instance oscillatory boundary conditions
\ba
 E(0,t)&=& E_0 e^{-i\beta t}
      \ =\ \frac{E_0}{2\pi i}
            \int\limits_{\sigma_1-i\infty}^{\sigma_1+i\infty}
             \frac{e^{p_1 t} dp_1}{p_1+i\beta}\ ,
\label{eq-5}\\
 f^{(1)}(0,t,\vec v)
       &=& \alpha e^{-i\beta t}
      \ =\ \frac{\alpha}{2\pi i}
            \int\limits_{\sigma_1-i\infty}^{\sigma_1+i\infty}
             \frac{e^{p_1 t} dp_1}{p_1+i\beta}\ ,
\label{eq-6}
\ea
one obtains for electrons/ions:
\ba
 f^{(1e)}_{p_1p_2}
   &=& \frac{1}{p_1+v_xp_2}
        \left[-\frac{v_x|e|}{kT_e}
         \left(\frac{m}{2\pi kT_e}\right)^{3/2}
          e^{-\frac{mv^2}{2kT_e}}E_{p_1p_2}
    + \frac{v_x\alpha_e\left(\vec v\right)}{p_1+i\beta}
    + f^{(1e)}_{p_2}\left(\vec v\right)\right]
\label{eq-7}\\
 f^{(1i)}_{p_1p_2}
   &=& \frac{1}{p_1+v_xp_2}
        \left[\frac{v_x|e|}{kT}
         \left(\frac{M}{2\pi kT}\right)^{3/2}
          e^{-\frac{Mv^2}{2kT}}E_{p_1p_2}
    + \frac{v_x\alpha_i\left(\vec v\right)}{p_1+i\beta}
    + f^{(1i)}_{p_2}\left(\vec v\right)\right]
\label{eq-8}
\ea
where Maxwell functions have been used:
\be
 f^{(e)}_0\left(\vec v\right)
  = \left(\frac{m}{2\pi kT_e}\right)^{3/2}
     e^{-\frac{mv^2}{2kT_e}}\ ;\quad
 f^{(i)}_0\left(\vec v\right)
  = \left(\frac{M}{2\pi kT}\right)^{3/2}
     e^{-\frac{Mv^2}{2kT}}\ ,
\label{eq-9}
\ee
and $f^{(1)}_{p_1}\left(\vec v\right)$,
$f^{(1)}_{p_2}\left(\vec v\right)$ are corresponding
{\em one-dimensional} (either in $t$ or $x$) Laplace images
of boundary and initial conditions (see~(\ref{eq-4})).
Then equation for finding $E_{p_1p_2}$ can be
written in the following form:
\ba
  E_{p_1p_2}
   \left[p_2
         -\frac{4\pi e^2n_e}{kT_e}
          \left(\frac{m}{2\pi kT_e}\right)^{3/2}
          \int\limits_{-\infty}^{\infty}
          \frac{e^{-\frac{mv^2}{2kT_e}}\cdot v_x d^3\vec v}
               {p_1+v_xp_2}
         -\frac{4\pi e^2n_e}{kT}
          \left(\frac{M}{2\pi kT}\right)^{3/2}
          \int\limits_{-\infty}^{\infty}
          \frac{e^{-\frac{Mv^2}{2kT}}\cdot v_x d^3\vec v}
               {p_1+v_xp_2} \right]\ =
\nonumber \\
 =\ \frac{E_0}{p_1+i\beta}
  - 4\pi|e|n_e\left[
      \int\limits_{-\infty}^{\infty}
       \frac{\left(\alpha_e-\alpha_i\right)\left(\vec v\right)
              v_x d^3\vec v}
             {\left(p_1+v_xp_2\right)\left(p_1+i\beta\right)}
      +\int\limits_{-\infty}^{\infty}
        \frac{\left(f^{(1e)}_{p_2}-f^{(1i)}_{p_2}\right)d^3\vec v}
             {\left(p_1+v_xp_2\right)}
         \right] .
\label{eq-10}
\ea

   We can transform integrals in the l.h.s. of eq.(\ref{eq-10}):
\be
 \int\limits_{-\infty}^{\infty}
  \frac{e^{-\frac{mv^2}{2kT_e}}\cdot v_x d^3\vec v}
       {p_1+v_xp_2}
 \equiv -2p_2
 \int\limits_{0}^{\infty}
  \frac{e^{-\frac{mv^2}{2kT_e}}\cdot v_x^2 d^3\vec v}
       {p_1^2-v_x^2p_2^2}
 \simeq -\left(\frac{2\pi kT_e}{m}\right)^{3/2}
            \frac{p_2}{2}\
             \frac{\bar{v_x^2}}{p_1^2-\bar{v_x^2}p_2^2} ,
\label{eq-11}
\ee
where $\bar{v_x^2}$ is some characteristic value of squared velocity
defined by Maxwellian exponent and $\gamma\simeq 1$
(see also a discussion on principal value sense of such integrals
in~\cite{bib-4}).
Note that the points $p_1=\mp v_x p_2$ here are not really poles!
The infinities are cancelled after account for the analogous
terms with $p_1\pm v_x p_2$ in denominator 
from $f_{p_2}^{(1)}(\vec v)$, with latter being proportional 
to $E_{p_1p_2}$ (see~\cite{bib-3,bib-4}).
Let us for simplicity omit off the consideration for a time
the additive terms proportional to
$f^{(1e)}\left(0,t,\vec v\right)$, $f^{(1i)}\left(0,t,\vec v\right)$,
$f^{(1e)}\left(x,0,\vec v\right)$ 
and $f^{(1i)}\left(x,0,\vec v\right)$ and obtain
\be
 E_{p_1p_2} = \frac{\frac{E_0}{p_2\left(p_1+i\beta\right)}}
                   {1
                   +\frac{\omega_e^2}{p_1^2-ap_2^2}
                   +\frac{\omega_i^2}{p_1^2-bp_2^2}}\ ,
\label{eq-12}
\ee
where $\omega_e$, $\omega_i$ are correspondingly Langmuir
electron and ion frequencies;
$\sqrt{a}=\sqrt{2\gamma kT_e/m}$, $\sqrt{b}=\sqrt{2\gamma kT/M}$
are nearby thermal mean velocities of one-dimensional thermal 
movement of electrons and ions.
\footnote{It is necessary to correct in~\cite{{bib-3},{bib-4}}
the term $E_0/p_2^2$ by the term $E_0/p_2$.
In Eq.(28) the second pair of poles must be
$\left(p_1=\pm i\bar\xi\sqrt{2Ae^{-\bar\xi^2}\Delta\xi},
       p_2=0 \right)$}
The value $\gamma$ can be approached in concrete calculations
by means of an iteration process with more and more precise values
of the poles ($p_1$,$p_2$) of expression (\ref{eq-12}) with their
subsequent substitution in eqn.(\ref{eq-11}).
The equation for finding a pair of poles of the function 
$E_{p_1p_2}$ (\ref{eq-12}) we call ``general dispersion equation''
for given boundary and initial problem.

   If the terms $\alpha_e\left(\vec v\right)$,
$\alpha_i\left(\vec v\right)$ and $f^{(1e)}_{p_2}$,
$f^{(1i)}_{p_2}$ are non-zero, they can be presented as sums
of symmetrical and antisymmetrical (with respect to $v_x \to -v_x$)
parts and be reduced by analogous way to integrals
over interval $0\le v_x<\infty$ with the integrand denominator 
$\left(p_1^2-\bar{v_x^2}p_2^2\right)$.
This might lead also to an appearance of new plasma modes.
The plasma boundary oscillations apparently can be realized
by supplying periodic electrical potential to wide flat electrode
grid immersed into homogeneous plasma volume.

   Asymptotical both in $x$ and $t$ plasma modes
$\left(\omega^{(n)}, k^{(n)}\right)$ are defined by
a pair of residues (in $p_1$ and $p_2$) at poles of
function (\ref{eq-12}).
The case $p_2=0$ (that is, relative displacements of electron 
and ion parts as a whole) leads to the pole
$p_1=\pm i\sqrt{\omega_e^2+\omega_i^2}$.
It corresponds to the eigen-mode of non-damping
plasma oscillations with the frequency (but not residue)
independent on $\beta$.
To excite this mode the boundary field should be increased
fast enough up to the value $E_0$.

  Besides that there is a mode $p_2=0$, $p_1=-i\beta$,
which corresponds to forced plasma oscillations of electrons
and ions as a whole with the frequency of excitation source.

  The more interesting case is $p_1=-i\beta$ and $p_2$ being the
root of equation
\be
 1 - \frac{\omega_e^2}{\beta^2+ap_2^2}
   - \frac{\omega_i^2}{\beta^2+bp_2^2}
   = 0\ .
\label{eq-13}
\ee
Its solution can be found analytically in general form:
\be
 p_2^2 = \frac{-1}{2ab}
          \left[a\left(\beta^2-\omega_i^2\right)
               +b\left(\beta^2-\omega_e^2\right)\right]
      \pm\frac{1}{2ab}
          \sqrt{\left[a\left(\beta^2-\omega_i^2\right)
                     +b\left(\beta^2-\omega_e^2\right)\right]^2
            -4ab\beta^2\left(\beta^2-\omega_e^2-\omega_i^2\right)}.
\label{eq-14}
\ee

   We do not consider here all particular cases, but instead note
that at common conditions $M \gg m$, $T_e \gg T_i$
with $\beta \gg \omega_i$ one obtains the following roots:
\be
  p_2^2
   \simeq -\frac{\beta^2-\omega_i^2}{b}\ ; \quad
  p_2^{(1,2)}
   \simeq \pm i\beta\sqrt{\frac{1-\omega_i^2/\beta^2}{b}}\ ,
\label{eq-15}
\ee
so the wave speeds (if there are no standing waves) are
\be
 v^{(1,2)}
   \simeq \pm\sqrt{\frac{b}{1-\omega_i^2/\beta^2}}
   = \pm\sqrt{\frac{2\gamma kT}{M}}\cdot
      \frac{1}{\sqrt{1-\omega_i^2/\beta^2}}
\label{eq-16}
\ee
-- that is the modes are the Langmuir ion waves.

  If $\beta\ll \omega_i$, the expression under the root sign
can be expanded in the small parameter
$$\frac{4ab\beta^2\left(\beta^2-\omega_e^2-\omega_i^2\right)}
       {\left[a\left(\beta^2-\omega_i^2\right)
             +b\left(\beta^2-\omega_e^2\right)\right]^2}\ .$$
At $\left[a\left(\beta^2-\omega_i^2\right)
         +b\left(\beta^2-\omega_e^2\right)\right] < 0$
one of solutions will take the form
\be
  p_2^2
   \simeq \frac{-\beta^2\omega_e^2}
               {a\omega_i^2+b\omega_e^2}\ ; \quad
  p_2^{(3,4)}
   \simeq \pm \frac{i\beta\omega_e}
               {\sqrt{a}\omega_i
                \sqrt{1+b\omega_e^2/a\omega_i^2}}\ ,
\label{eq-17}
\ee
so the travel speeds are
\be
 v^{(3,4)}
   \simeq \pm\frac{\sqrt{a}\omega_i}{\omega_e}
           \sqrt{1+b\omega_e^2/a\omega_i^2}
   = \pm\sqrt{\frac{2\gamma kT_e}{M}
              \left(1+\frac{T}{T_e}\right)}\ ,
\label{eq-18}
\ee
that is these solutions correspond to the modes
of non-damping ion-acoustic waves.
The other pair of values $p_2$ are
\ba
 p_2^2
  &\simeq&
   \frac{1}{ab}
   \left|a\left(\beta^2-\omega_i^2\right)
        +b\left(\beta^2-\omega_e^2\right)\right|
   \simeq\frac{\omega_i^2-\beta^2}{b}\ ;
\nonumber \\
 p_2^{(1,2)}
  &\simeq& \pm\omega_i
            \sqrt{\frac{1-\beta^2/\omega_i^2}{b}}
\label{eq-19}
\ea
and correspond to exponential damping and exponential
growing modes.

  Besides that in all the cases the damping solutions
are not of the specific Landau damping type.

   As it has been noted in~\cite{bib-4},
the presence of growing solutions is connected with
the inconsistency of the initial and boundary conditions
on the field $E(0,t)$ and functions $f^{(1)}\left(0,t,\vec v\right)$
and $f^{(1)}\left(x,0,\vec v\right)$ , which are not independent.
The initial and boundary conditions must be given in such a way,
that the residues sums in the coefficients of
exponentially growing modes be cancelled.
Such consistency condition leads to linear equations
connecting pre-exponential coefficients of different modes
and makes the existence of plasma oscillations
not independent on electric field impossible.

  Laplace images of the form (\ref{eq-12}) or (\ref{eq-20})
are general solutions of given concrete problems after
applying inverse Laplace transformation. 
It stays now but an open question, whether one can
with integration over $p_2$, say, to pass to
1-dimensional Landau initial problem and
to his 1-dimensional Laplace image. This is in
accord with above statement that any plasma oscillatory
problem is not strongly only initial or only boundary, 
but instead mixed problem with coupled oscillatory modes.

  It should be noted that a single-valued solution
of Vlasov first-order partial differential equations
for the self-consistent plasma electric field $E_{con}(x,t)$
is completely defined by initial condition $E_{con}(x,0)$
and by a single-point boundary condition $E_{con}(0,t)$ only.
It appears to contradict to a possibility of an arbitrary setting
electric field in several points using some electrode system.
This contradiction is removed by taking into account
two different parts of the total electric field
$E(x,t)=E_{con}(x,t)+E_{ext}(x,t)$ in 
eqns.(\ref{eq-1a}-\ref{eq-1b}),
where $E_{ext}(x,t)$ is inconsistent {\em external} field
(its inconsistency means that it is defined by external source,
not by plasma).
In practice one usually transits from field to potential
according to $E_{ext}=-\dpa\phi_{ext}/\dpa x$,
and $E_{con}$ must satisfy Poisson equation (\ref{eq-2})
and is {\em defined} as self-consistent field
with boundary condition $E_{con}(0,t)$.

\section{Proper electron waves in the low-collision plasma}

  Adding into the r.h.s. of eqn.(\ref{eq-1a}) the collision term
$-\nu\left(\bar v\right)f^{(1)}$ in close analogy
with the previous section we obtain
\be
 E_{p_1p_2} = \frac{\frac{E_0}{p_2\left(p_1+i\beta\right)}}
                   {1+\frac{\omega_e^2}
                           {\left(p_1+\nu\right)^2-ap_2^2}}\ ,
\label{eq-20}
\ee
where $\nu\left(\bar v\right)$ is an effective collision frequency
between electrons and neutral particles.
At $p_2=0$ one obtains the pole $p_1=\pm i\omega_e-\nu$
-- that means the damping in time Langmuir oscillations.
Besides that there are non-damping oscillations
with the boundary field frequency ($p_2=0$, $p_1=-i\beta$).

   At $p_1=-i\beta$ one obtains also according to (\ref{eq-20})
the equation for determining $p_2$:
\be
  p_2^2
   = \frac1{a}\left[\left(\nu^2-\beta^2+\omega_e^2\right)
                   -2i\beta\nu\right]\ .
\label{eq-21}
\ee
Omitting the elementary algebraic procedures of
extracting the root square of a complex variable,
we give the final result:
\be
 p_2
  = \frac{\pm\beta\nu\sqrt{\frac2{a}}}
         {\left(\sqrt{F^2+4\beta^2\nu^2}-F\right)^{1/2}}
  \mp\frac{i}{\sqrt{2a}}
          \left(\sqrt{F^2+4\beta^2\nu^2}-F\right)^{1/2}\ ,
\label{eq-22}
\ee
where $$F \equiv \nu^2-\beta^2+\omega_e^2\ .$$

   At $F<0$ and small $\nu\ll\sqrt{\beta^2-\omega_e^2}$,
$\beta>\omega_e$ one obtains
\be
 p_2
  \simeq \frac{\pm\beta\nu\sqrt{a}}
         {\sqrt{\beta^2-\omega_e^2}}
  \mp\frac{i}{\sqrt{a}}
          \sqrt{\beta^2-\omega_e^2}\ .
\label{eq-23}
\ee

   At $F>0$ it is more convenient to rewrite eqn.(\ref{eq-22})
in the form
\be
 p_2 =
     \pm\frac{\left(\sqrt{F^2+4\beta^2\nu^2}+F\right)^{1/2}}
             {\sqrt{2a}}
     \mp i\frac{\beta\nu\sqrt{\frac2{a}}}
             {\left(\sqrt{F^2+4\beta^2\nu^2}+F\right)^{1/2}}\ ,
\label{eq-24}
\ee
and at $\nu \ll \sqrt{\omega_e^2-\beta^2}$, $\omega_e>\beta$
one obtains
\be
 p_2
  \simeq \pm\frac{\sqrt{\omega_e^2-\beta^2}}
                 {\sqrt{a}}
         \mp\frac{i\beta\nu}{\sqrt{a}}
         \frac{1}{\sqrt{\omega_e^2-\beta^2}}\ .
\label{eq-25}
\ee

  In both cases one obtains the exponential growing solutions.
But substitution of any exponential growing expression
of the form $\exp(i\omega t + ikx)$ with complex values
$\omega$, $k$ in eqns.(\ref{eq-1a})--(\ref{eq-2})
leads to a dispersion equation which has no solutions~\cite{bib-4}.
It means that at the same extent as
taking into account the boundary condition $E(0,t)$
one must also for the sake of consistency with $E(0,t)$
take into account initial and boundary values of functions
$f^{(1e)}\left(x,t,\vec v\right)$,
$f^{(1i)}\left(x,t,\vec v\right)$, in such a way
that growing terms be cancelled
(as it has been already noted in~\cite{bib-4}).

  The possibility of such cancellation is provided with
the same poles of growing solutions for $E_{p_1p_2}$
and Laplace images of initial and boundary values
of $f^{(1e)}$ and $f^{(1i)}$.

\section{Conclusions}

  We have considered Vlasov differential equations
for collisionless plasma and successfully solved them
asymptotically by the Laplace transform method
of operational calculus.

  We have proposed absolutely new, very simple and effective way
of finding plasma oscillation modes: they are defined
by the pairs ($\omega^{(n)}$, $k^{(n)}$), $n = 1, 2, \ldots$,
which are determined as pairs of roots (``double-zeros'')
of the ``generalized dispersion equation''
\be
 \frac{1}{E_{p_1p_2}} = 0\ ,
\label{eq-26}
\ee
where $E_{p_1p_2}$ is two-dimensional (in $x$ and $t$)
Laplace image of self-consistent plasma electric field $E(x,t)$.
Some additional plasma density oscillation modes appear
to be determined from
\be
 \frac{1}{f_{p_1p_2}^{(1e)}} = 0\ ;\quad
 \frac{1}{f_{p_1p_2}^{(1i)}} = 0\ ,
\label{eq-27}
\ee
but really this has no place.
According to the required mutual consistency
of initial and boundary values of $f^{(1e)}\left(x,t,\vec v\right)$,
$f^{(1i)}\left(x,t,\vec v\right)$ with boundary field $E(0,t)$
(the solutions finiteness condition)
these additional oscillatory modes must be connected with
the electric field, thus the coefficients in the total sum of 
the modes have to be proportional to $E_0$.

  The eqn.(\ref{eq-26}) is in principle different from
equation commonly used in literature (see, for instance,
the classic text-book exposition in~\cite{{bib-7},{bib-8}})
\be
  \varepsilon_l\left(\omega,k\right) = 0\ ,
\label{eq-28}
\ee
where (according to the considered in this paper cases)
$\varepsilon_l$ is longitudinal dielectric susceptibility
defined by only the intrinsic parameters
of a homogeneous infinite plasma.
Contrary to this approach the function $E_{p_1p_2}$
accounts for initial and boundary conditions concretely.
The finding $E_{p_1p_2}$, contrary to $\varepsilon_l$,
does not require using Landau theory with its doubtful
foundations (using analytical continuation,
expansions in asymptotically divergent series 
(see~\cite{bib-4,bib-6}),
problems with the principle of causality and so on).
At the same time the equation (\ref{eq-26}) allows one,
contrary to (\ref{eq-28}), to find {\em all}
the oscillation modes in a given concrete problem.
It reveals immediately the tight interconnection of
oscillatory modes with the concrete conditions,
that is with methods of excitation, excitation frequency,
functional forms of initial and boundary conditions,
whereas eqn.(\ref{eq-28}) connects plasma oscillations and waves
only with proper intrinsic parameters of plasma.

  Moreover, the required connection of initial and boundary
conditions means that it is impossible completely
to discriminate between a pure boundary or pure initial problems.
So, to find the finite solution of a ``boundary'' problem
it is certainly required to supply also non-zero initial values
of distribution functions $f^{(1)}$ under consideration.

  It should be emphasized once again that in general case
using the theory of plasma oscillations, based on 
susceptibility $\epsilon$ and plasma modes equation $\epsilon=0$,
either is erroneous (because it is based on the dispersion
equation which does not exist) or has rather limited 
range of applicability.

  In this paper we have considered only the simplest examples
of application of suggested method.
Besides of diverse generalizations one of the main problems
here stays the account for Coulomb collisions,
which requires cumbersome mathematical computations
with revision of results described in literature
(see, for instance,~\cite{bib-6}) and based mainly
on expansions in asymptotically divergent series
in small parameter $\delta/\omega_0$.

\acknowledgements

  My sincere thanks to Prof.~A.~A.~Rukhadze for his periodic
supports and at the same time continued distrustful criticism.
All my thanks also to Dr.~A.~P.~Bakulev for a fruitful and
very constructive and critical discussion
and a help in getting this paper up.

\end{document}